# Multi-focal laser surgery: cutting enhancement by hydrodynamic interactions between cavitation bubbles

I. Toytman[1]*, A. Silbergleit[1], D. Simanovski[1], and D. Palanker[1,2]

[1]*Hansen Experimental Physics Laboratory, Stanford University, 452 Lomita Mall, Stanford, CA 94305 USA*
[2]*Department of Ophthalmology, Stanford University School of Medicine, 300 Pasteur Drive, Stanford, CA 94305 USA*

Transparent biological tissues can be precisely dissected with ultrafast lasers using optical breakdown in the tight focal zone. Typically, tissues are cut by sequential application of pulses, each of which produces a single cavitation bubble. We investigate the hydrodynamic interactions between simultaneous cavitation bubbles originating from multiple laser foci. Simultaneous expansion and collapse of cavitation bubbles can enhance the cutting efficiency by increasing the resulting deformations in tissue, and the associated rupture zone. An analytical model of the flow induced by the bubbles is presented and experimentally verified. The threshold strain of the material rupture is measured in a model tissue. Using the computational model and the experimental value of the threshold strain one can compute the shape of the rupture zone in tissue resulting from application of multiple bubbles. With the threshold strain of 0.7 two simultaneous bubbles produce a continuous cut when applied at the distance 1.35 times greater than that required in sequential approach. Simultaneous focusing of the laser in multiple spots along the line of intended cut can extend this ratio to 1.7. Counter-propagating jets forming during collapse of two bubbles in materials with low viscosity can further extend the cutting zone – up to a factor of 1.54.

## I. INTRODUCTION.

Ultrafast lasers are used in a wide variety of biomedical applications. The most common surgical application is creating a corneal flap in refractive surgery (a part of LASIK procedure), replacing less precise mechanical microkeratome. Femtosecond lasers are also used in keratoplasty [1], where they allow for custom shaping of the corneal transplant, which facilitates postoperative wound healing. Applicability of ultrafast lasers has been recently demonstrated in cataract surgery [2]. An anterior lens capsule was cut in a spiral pattern forming a continuous cylindrical surface. Lens cortex was dissected into segments by cutting along several planes forming a cross and various boxed patterns. Another promising application of ultrafast lasers in ophthalmology is the crystalline lens softening by patterned incisions within the peripheral lens cortex. Such procedure was shown to increase the lens flexibility and therefore could be used for the presbyopia treatment [3]. Several neurological applications of ultrafast lasers have been demonstrated, including single neuron dissection in a living organism [4] and excision of brain tissue from histological preparations [5]. Various subcellular structures such as chromosomes, mitochondria, plant cell walls and single chloroplasts [6-8] can be cut with femtosecond pulses while maintaining cell viability. Successful cell transfection involving membrane poration with an ultrafast laser has also been performed [9].

All these applications utilize optical breakdown inside transparent tissue produced by a tightly focused short-pulsed laser beam. Ultra-short pulse duration allows achieving sufficiently high peak power for multiphoton ionization with relatively low pulse energies. For sub-picosecond pulses, the threshold fluence for ionization in transparent tissue is on the order of 1 $J/cm^2$ [10]. For the beam focused in a micrometer diameter spot this corresponds to a few nanojoule pulse energy. During optical breakdown a fraction of the pulse energy is absorbed in the focal volume leading to explosive vaporization of the interstitial water. Expansion of the vapor bubble can mechanically rupture the surrounding material at distances much larger than the size of the focal zone. Reduction of the threshold energy by shortening pulse duration and tightening focal spot helps minimizing the deposited energy and thus reduces the collateral damage zone produced by cavitation bubble.

* Corresponding author: itoytman@stanford.edu

A continuous cut is typically produced by sequential application of laser pulses, placed sufficiently close, so that the individual rupture zones coalesce. In ophthalmic applications, the constant eye movements that persist even with the attached suction ring impede precise positioning of pulses. Exact placement of the laser focus is also difficult in dissection of poorly visible structures, such as a cell membrane or an individual nerve axon. In such applications a series of pulses may be applied to produce a linear cut rather than a single pulse. However, tissue movement induced by the first pulse of the series may make its subsequent targeting more difficult. These problems can be avoided, in principle, if multiple cavitation bubbles are produced at once, with a target tissue trapped between them. In this paper we analyze hydrodynamic interactions between two simultaneously created bubbles, and compare the resulting rupture zone to that produced in a sequential approach.

## II. THEORETICAL FORMULATION

Expansion of a cavitation bubble within a sample creates deformations that can be characterized by the strain tensor $\varepsilon$. Deformation of an infinitesimally small volume can be decomposed into three independent strains (corresponding to either stretching or compression) along mutually orthogonal axes, known as principle axes. In this representation the diagonal components of $\varepsilon$ are the tensor eigenvalues (principle strains), and the non-diagonal ones are zero. We assume that the material is ruptured if at least one of the principle strains exceeds certain threshold $\varepsilon_{th}$, which is considered to be an intrinsic material property. In the rest of the paper we will be using the term "strain" for the largest of the three principle strains.

In the simplest case of a single spherical bubble of the initial radius $R_1$ expanding within incompressible material to some final radius $R_2$, the material enclosed between the spheres with the radii $R_1$ and $r_1>R_1$ will be transformed into a layer enclosed by spheres with radii $R_2$ and $r_2$. Using volume conservation condition we can find $r_2$:

$$r_2 = \sqrt[3]{r_1^3 + R_2^3 - R_1^3}\,, \tag{1}$$

and then the strain:

$$\varepsilon = \frac{r_2 - r_1}{r_1} = \left[1 + \frac{R_2^3 - R_1^3}{r_1^3}\right]^{1/3} - 1\,. \tag{2}$$

For a certain threshold strain $\varepsilon_{th}$ the radius of the rupture zone $R_{rupture}$ is then:

$$R_{rupture} = \sqrt[3]{\frac{R_2^3 - R_1^3}{(1+\varepsilon_{th})^3 - 1}} \approx \frac{R_2}{\sqrt[3]{(1+\varepsilon_{th})^3 - 1}}\,, \tag{3}$$

where the last equality assumes $R_1 << R_2$.

In the case of multiple bubbles the displacements in the material can be determined by solving the proper boundary value problem. We assume adiabatic expansion of two cavitation bubbles in an inviscid and incompressible fluid. These assumptions have been shown to describe the behavior of a single cavitation bubble in water very accurately [11]. The resulting flow is defined by its potential $\Phi$, obeying the Laplace equation:

$$\Delta\Phi = 0 \tag{4}$$

and the following boundary conditions:

$$\left.\frac{\partial\Phi}{\partial\vec{n}}\right|_S = v_b\,, \qquad P_{gas} = p_b\,. \tag{5}$$

Here $S$ and $v_b$ represent the boundary surface and its velocity for each of the bubbles, $P_{gas}$ is the gas pressure inside the bubble, and $p_b$ is the pressure of the liquid at the bubble boundary. The latter is expressed through the potential using the Bernoulli integral of Euler equations:

$$\rho\left[\frac{\partial\Phi}{\partial t} + \frac{1}{2}(\nabla\Phi)^2\right]\Bigg|_S + p_b = p_\infty\,, \tag{6}$$

where $p_\infty$ is the hydrostatic pressure and $\rho$ is the liquid density.

In the case of a single bubble this problem was solved by Rayleigh [12]. Though his derivation assumes $P_{gas}=const$, it is fairly straightforward to generalize the solution for any function $P_{gas}(V)$ (see Appendix A). The result is that the flow potential $\Phi_{single}$ in this case is just a monopole ("charge"),

$$\Phi_{single}(r,t) = \frac{A(t)}{r} \; , \tag{7}$$

whose strength, $A$, changes with the time according to the formula

$$A(t) = -\left( \frac{2}{3} V^{1/3} \int_{V_0}^{V(t)} g(v)dv \right)^{1/2} . \tag{8}$$

Here $V(t)$ is, up to a factor, the bubble volume $V(t)=a^3(t)$, and $a(t)$ is the current bubble radius (determined in Appendix A). The normalized difference between the pressure inside the bubble and far away from it that drives the bubble dynamics

$$g(V) = \frac{p_\infty}{\rho}\left( \frac{p_0 V_0^{5/3}}{p_\infty V^{5/3}} - 1 \right). \tag{9}$$

(note that we use the adiabatic index $\gamma=5/3$); finally, $p_0$ is the initial gas pressure inside the bubble, and $V_0$ is the initial value of $V(t)$.

The case of the two initially spherical bubbles is essentially more complicated, since their shapes change with time in an unknown way, and one faces thus a non-linear problem with an unknown boundary. To find its solution, we develop here an analytical perturbative approach, representing the flow potential as a superposition of axially symmetric multi-poles in two spherical coordinate systems with the origins at the centers of the bubbles, and simultaneously expanding the unknown for $t>0$ shapes of their boundaries as functions of the corresponding polar angles (Fig. 1).

Extensive numerical studies of the bubble dynamics used the boundary element method [13], mostly in applications to interactions with a plane boundary (see [14 – 16] and the references therein), and in conjunction with a recent analytical model for 2-dimensional bubble geometry [17]. If the boundary is rigid, the problem is then equivalent to that of two identical bubbles. The evolution of the bubble shapes during both their growth and collapse was studied in detail, but the flow potential is usually not given explicitly**,** and the strain needed for our study is not determined. The material properties (such as viscosity, elastic modulus and failure stress) were included in yet another numerical model of cavitation bubble dynamics [18] for the case of a single spherical bubble; this model was later extended to 2D cylindrical geometry, in which a bubble was produced on a fiber tip [19, 20]. Our analytical approach, being not as powerful as the numerical methods described in terms of strongly non-linear regimes of the bubble interaction, has its own advantages. First of all, it is a direct and rather simple *analytical* generalization of the Rayleigh solution enabling various physical insights into the picture of the flow and scaling with various parameters. Second, it works for two bubbles of different sizes, and captures an effect of a uniform drift of the bubble system through the fluid with just the first correction (this effect is missing in the case of identical bubbles, see below). It also allows one to get the bubble shapes in a good agreement with the experiment (as well as in a qualitative agreement with the results of previous studies [14 – 16]), and to easily compute the needed strain distribution.

So, we expand both the potential $\Phi(r, \theta, t)$ and the function of boundary shape $R(\theta, t)$ in a series of successive corrections,

$$\Phi(\vec{r},t) = \Phi^{(0)}(\vec{r},t) + \Phi^{(1)}(\vec{r},t) + \Phi^{(2)}(\vec{r},t) + \Phi^{(3)}(\vec{r},t) + \ldots \tag{10}$$

$$R(\theta,t) = a(t) + f(\theta,t) = a(t) + f^{(1)}(\theta,t) + f^{(2)}(\theta,t) + f^{(3)}(\theta,t) + \ldots \tag{11}$$

where $\Phi^{(0)}$ is a superposition of two monopoles, each giving the exact solution for either of the two bubbles in the absence of the other one, and $R(\theta, t)$ describes the shape of each bubble. In these approximations we assume $|f/a| << 1$ and $|f^{(i+1)}/f^{(i)}| << 1$. We denote $L$ the initial distance between bubble centers.

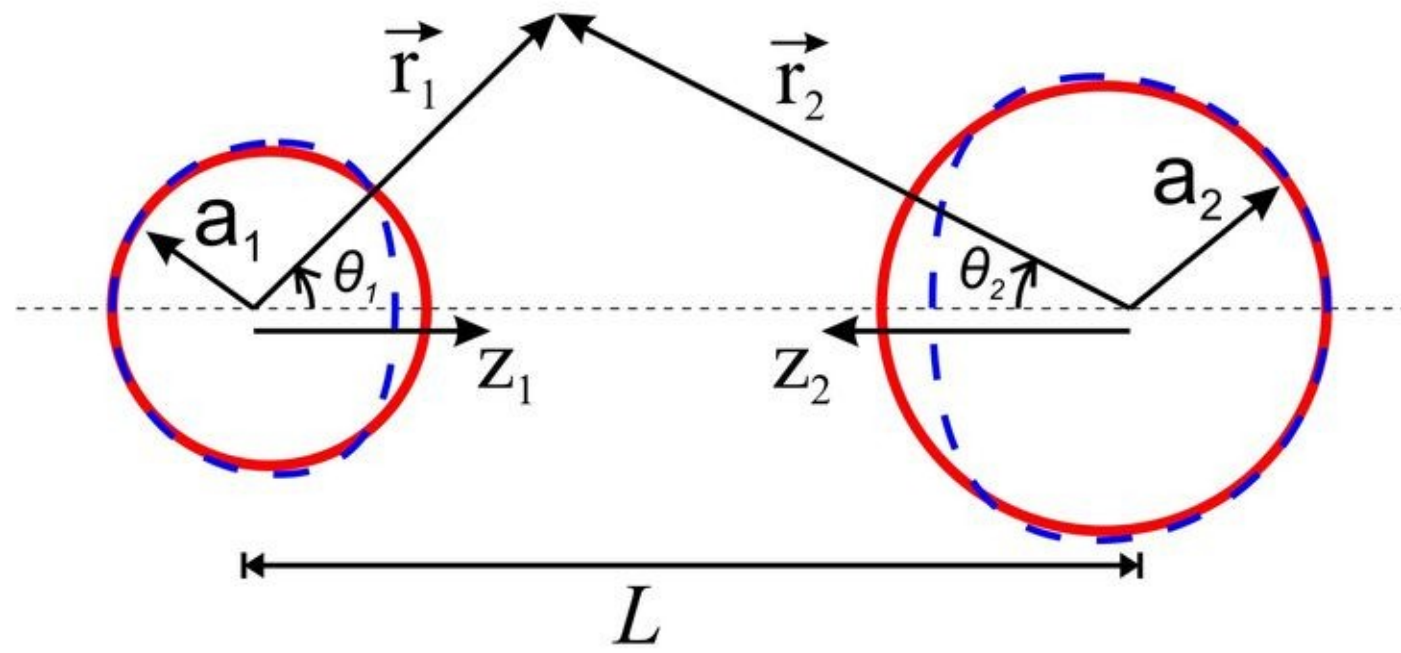

FIG. 1. (Color online) Model geometry in general case of non-identical bubbles. Dashed and solid lines illustrate bubble shape with and without presence of the other bubble, respectively.

It can be shown (see Appendix B) that there is no first-order correction to the bubble shape ($f^{(1)}\equiv 0$) and the first order addition to the potential is:

$$\Phi^{(1)} = -A_1(t)\frac{r_1\cos\theta_1}{L^2} - A_2(t)\frac{r_2\cos\theta_2}{L^2} = -A_1(t)\frac{z_1}{L^2} - A_2(t)\frac{z_2}{L^2}. \quad (12)$$

Here and in the rest of the paper indices 1 and 2 refer to the coordinate systems associated with the first and the second bubble, respectively. Note that in the case of bubbles with different initial radii the functions $A_1(t)$ and $A_2(t)$ are not identical, and the expression (12) describes the uniform flow along the z axis with the velocity proportional to the difference $A_1(t)-A_2(t)$. This flow corresponds to the drift of both bubbles, in agreement with the previous experimental results [21].

For the rest of the paper (except the Appendices B and C) we assume that the bubbles are identical and write all the equations in the system of the first bubble, thus dropping the index 1 wherever it does not cause any confusion. Note that the complete problem includes also the second set of equations (for the second bubble) that can be obtained by simply exchanging the indices 1 and 2. For the second order approximation to the boundary shape $f^{(2)}$ and the potential $\Phi^{(2)}$ we obtain (see Appendix C for the derivation):

$$f^{(2)}(\theta,t) = \alpha(t)\cos\theta; \quad (13)$$

$$\Phi^{(2)} = C(t)\left(\frac{\cos\theta_1}{r_1^2} + \frac{\cos\theta_2}{r_2^2}\right). \quad (14)$$

The functions $\alpha(t)$ and $C(t)$ can be found from the following system of ODE:

$$\dot{\alpha} = \frac{A}{L^2} + \frac{2\alpha A}{a^3} - \frac{2C}{a^3}; \quad (15)$$

$$\dot{C} = \alpha\dot{A} + \dot{\alpha}A - \frac{\dot{A}a^3}{L^2}. \quad (16)$$

In a similar way, the third order approximation can be written as:

$$f^{(3)}(\theta,t) = \beta(t)P_2(\cos\theta); \quad (17)$$

$$\Phi^{(3)} = D(t)\left(\frac{P_2(\cos\theta_1)}{r_1^3} + \frac{P_2(\cos\theta_2)}{r_2^3}\right). \quad (18)$$

where functions $\beta(t)$ and $D(t)$ describing the evolution are specified by the following differential equations:

$$\dot{\beta} = \frac{2Aa}{L^3} + \frac{2A\beta}{a^3} - \frac{3D}{a^4}; \quad (19)$$

$$\dot{D} = a(\dot{A}\beta + A\dot{\beta}) - \frac{\dot{A}a^5}{L^3}. \quad (20)$$

Knowing the flow velocity distribution at any given moment of time, the displacements and strain can be calculated either analytically or numerically. We have chosen and implemented the following numerical procedure. The entire bubble expansion time was divided into 1000 equal intervals *Δt*, and the flow within each interval was approximated by that in the beginning of the interval. The trajectory of any point could then be iteratively constructed by calculating the velocity *v* of the point and then adding the displacements *v*Δt* to the current coordinates of the point in each iteration.

To find the components $\varepsilon_{ij}$ of the strain tensor of an infinitesimal volume of material located prior to the bubble expansion at a point $\mathbf{r^{(0)}}$ with Cartesian coordinates $x_i^{(0)}$ (i=1,2,3) we first consider three "neighboring" points with coordinates of k-th point defined as:

$$x_i^{(k)} = x_i^{(0)} + d * \delta_{ik}\,, \tag{21}$$

where $d$ = *L/1000* and $\delta_{ik}$ is the Kronecker delta symbol. Let $X_i^{(k)}$ (k = 0,1,2,3) be the corresponding coordinates of $\mathbf{r^{(0)}}$ and its "neighboring" points after the bubble expansion. Replacing the derivatives by appropriate finite differences, the strain tensor components can be written as:

$$\begin{aligned}\varepsilon_{ij} &= \frac{1}{2}\left(\frac{\left(X_i^{(j)} - X_i^{(0)}\right) - \left(x_i^{(j)} - x_i^{(0)}\right)}{d} + \frac{\left(X_j^{(i)} - X_j^{(0)}\right) - \left(x_j^{(i)} - x_j^{(0)}\right)}{d}\right) \\ &= \frac{1}{2}\frac{\left(X_i^{(j)} - X_i^{(0)}\right) + \left(X_j^{(i)} - X_j^{(0)}\right)}{d} - \delta_{ij}\,.\end{aligned} \tag{22}$$

The proper strain at the final position corresponding to $\mathbf{r^{(0)}}$ is found as the largest of the three eigenvalues of the tensor $\varepsilon$. Boundary of the rupture zone is then determined as a surface where the strain reaches the threshold value $\varepsilon_{th}$.

## III. EXPERIMENTAL PROCEDURES

An amplified Ti:Sapphire laser system with the pulse duration of 1ps, 800nm wavelength, and the pulse energy up to 1mJ was used in the experiments. The system was operated at 10Hz repetition rate, and a single pulse could be selected with a mechanical shutter. The pulse energy was controlled by a variable attenuator consisting of a half-wave plate and a polarizer. To produce two cavitation bubbles simultaneously, the pulse was split by a thin-film polarizer (TFP) placed after the attenuator. To ensure equal energy in two resulting pulses another half-wave plate was introduced in front of the TFP.

Cavitation bubbles were imaged with fast flash photography, using a 10x microscope objective (Nikon, NA=0.3), a 250mm field lens, and a CCD camera (Photometrics CoolSnapHQ). The sample cuvette was made of thin (170µm) glass slides, and bubbles were generated near the wall to allow for high resolution imaging through a thin layer of material. The illumination was provided by an LED (OptoDiode Corp., OD-620L) focused onto a sample by a lens matching the NA of the imaging objective. Short pulses from a home-built pulse generator applied to the LED produced flash with duration as short as 100ns. Various moments of bubble dynamics were visualized by varying the delay between the laser pulse and the flash. The time-integrated image could be obtained by increasing the flash duration.

The threshold deformation and associated strain $\varepsilon_{th}$ was measured in gelatin with 90% water content (by mass), that served as a tissue model. Since the microcracks left after the collapse of the cavitation bubble could not be detected by our imaging system we designed a new method for visualization of the gelatin rupture. Solution of gelatin in water at 60°C was poured into the cuvette and a thin layer of oil was added on top of gelatin immediately afterwards to ensure flatness of the gelatin boundary and to serve as a contrast material facilitating visualization of the boundary. While visible due the difference in refractive indices, gelatin-oil interface produced relatively low distortion of hydrodynamic movement. In this experiment a single laser beam was focused by a 10x microscope objective (Olympus UMPlanFL, NA=0.3). Cavitation bubbles were created within the gelatin at various distances from the oil boundary. When this distance was smaller than the rupture zone, breakage of the gelatin boundary and ejection of gas into oil could be observed on the time-integrated images. Radius of the rupture zone was estimated as the maximum distance between the bubble center and the oil boundary, at which rupture occurred. The threshold strain, $\varepsilon_{th}$ can be derived from the measured radius of the rupture zone using Eq. 3.

For experimental verification of our theoretical model the bubbles were produced in distilled water by the laser pulses simultaneously applied in two distinct locations. The incident beams were focused into a cuvette using a 63x water-immersion microscope objective (Zeiss, NA=0.9), as shown in Fig. 2. Large NA helped avoiding self-focusing and produced highly spherical bubbles at well-defined locations. The distance between the centers of the bubbles, $L$ was adjusted by changing the angle $\varphi$ between the beams.

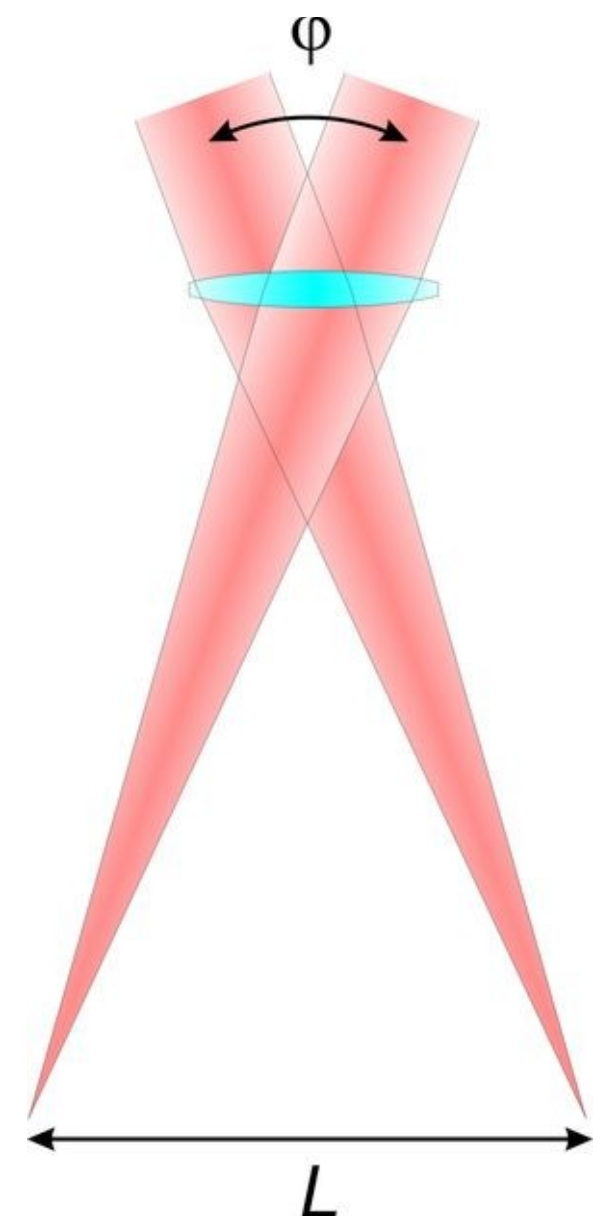

FIG. 2. (Color online) Focusing geometry for creating two identical simultaneous cavitation bubbles.

In contrast to water, deformations induced by cavitation bubbles in elastic and viscous materials (such as biological tissues) are difficult to compute analytically. However, in some cases these deformations can be visualized experimentally by embedding markers in the sample material. We tested this method in gelatin using 1μm polystyrene beads as position markers. The beads were added to gelatin solution during its preparation, thus their distribution in the sample volume was fairly uniform. The traces of the moving beads in tissue during expansion and collapse of the bubbles were imaged using an integrating exposure of 1ms in duration, which captured the entire process of the bubble evolution.

## IV. RESULTS AND DISCUSSIONS

### A. Threshold strain in gelatin

The maximum bubble radius observed in the time-integrated images in gelatin (Fig. 3) was ~24μm. A bubble created sufficiently far ($d$>40μm) from the gelatin-oil boundary deforms the boundary but does not rupture it (Fig. 3a). For $d$<40μm the boundary rupture and bubble penetration into oil is clearly visible as a sharp discontinuity (Fig. 3b,c). With $R_{rupture}$=24μm and $R_2$ = 40μm, we obtain from Eq. 3 $\varepsilon_{th}$ = 0.7.

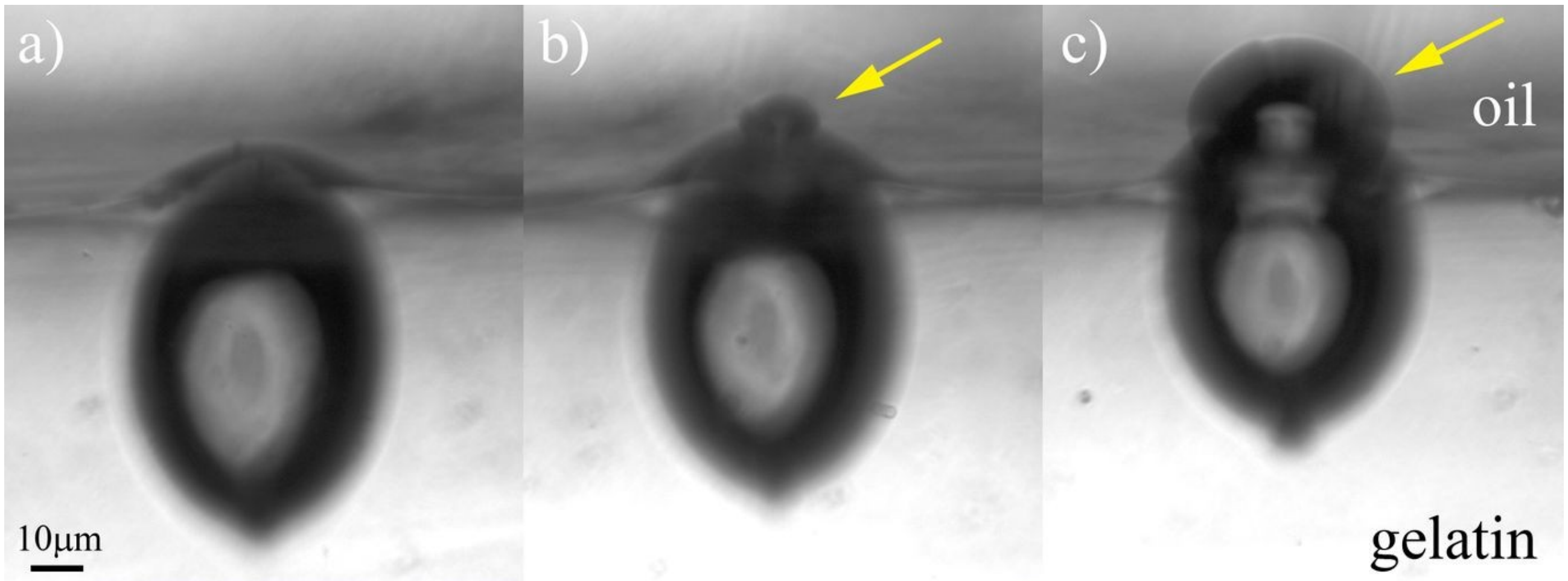

FIG. 3. (Color online) Time-integrated photographs of cavitation bubble in gelatin at various distances from the gelatin-oil boundary: a) no boundary rupture; b) minimal boundary rupture; c) profound boundary rupture. Yellow (light) arrows point to the bubble penetrating into the oil. Elongated shape of the bubbles is due to axially extended zone of optical breakdown owing to low NA of the objective and aberrations at multiple boundaries (glass-oil and oil-gelatin).

Published values of the threshold strain for the anterior lens capsule $\varepsilon_{th}$ range from 0.4 to 1.1 depending on the age, with higher values obtained in younger patients [22]. The value of 0.7, similar to our results with gelatin corresponds to the age 40-50. For human cornea the corresponding value is lower – only ~0.2 [23].

### B. Experimental verification of the bubble dynamics

To verify the validity of our theoretical model we compared the analytically derived bubble contours with those observed experimentally. As shown in Fig. 4 (top row) the exact analytical solution for a single bubble is in an excellent agreement with the experiment at every stage of bubble expansion, which validates the formulation and major assumptions of the model (negligible viscosity, and lack of heat exchange between the gas and the liquid). Remarkably, the solution for the two simultaneous bubbles, derived as an approximation for bubble sizes $R_{max}$ much smaller than the distance between the bubbles L, describes the bubble contours fairly well even for the ratio $R_{max}/L$~0.5 (Fig. 4, bottom row). The largest deviation of the theoretical contours from the experimental ones does not exceed ~20% and occurs at large angle θ~90°, while the results at smaller angles (θ<15°), which are of primary interest in this application, are even more accurate. It is worth noting that the discrepancy between theoretical and experimental bubble shapes is increasing with the bubble size and therefore is unlikely to be due to the neglect of surface tension, whose role decreases with the bubble growth.

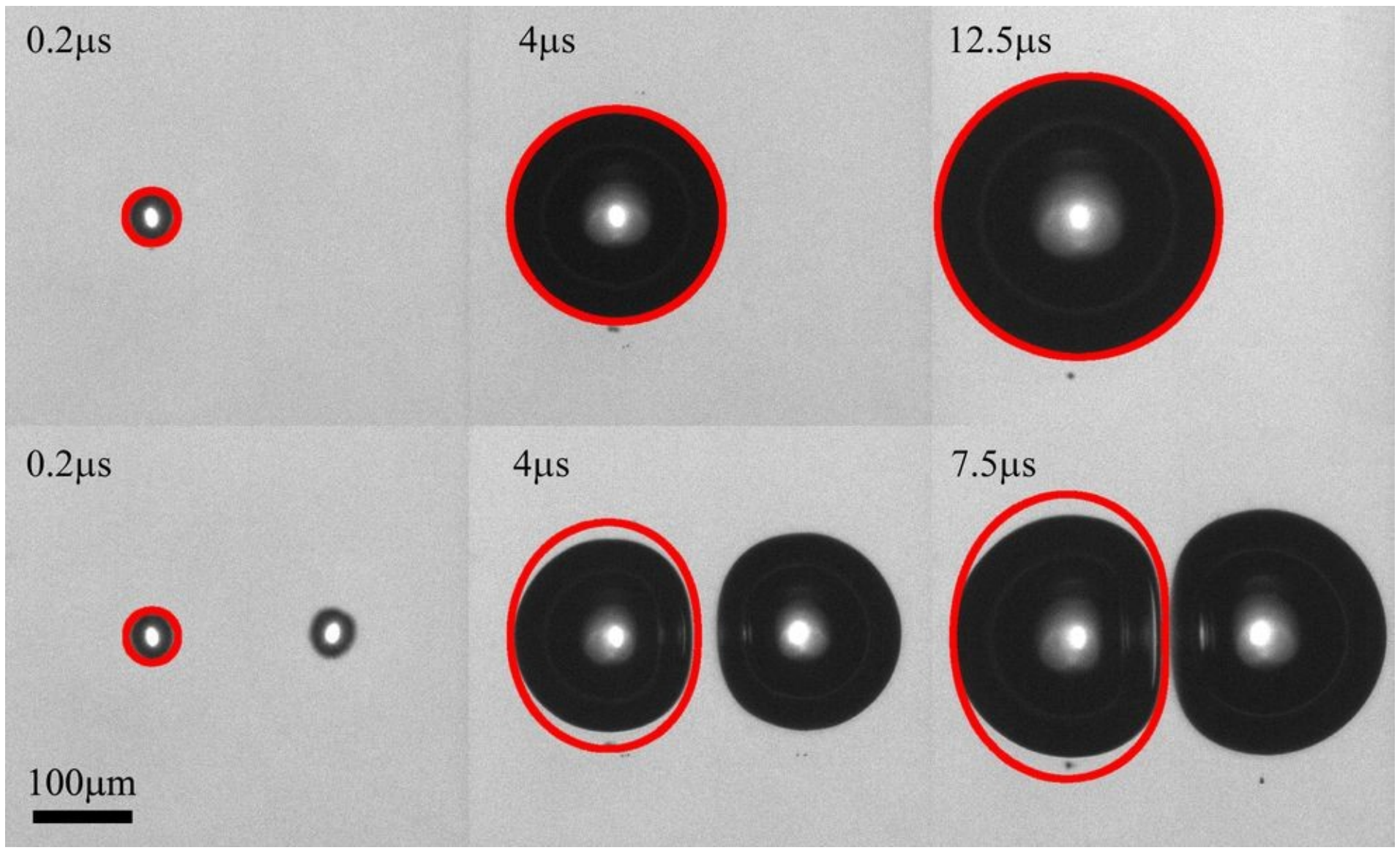

FIG. 4. (Color online) Analytically calculated bubble contours (solid lines) overlaid on the experimentally observed bubbles for a single bubble (top row) and for two simultaneous bubbles (bottom row).

As described above, the computation of strain based on the known flow involves certain choice of parameters such as the time step $\Delta t$ and the distance to the "neighboring" points. Comparison of the numerical and analytical (Eq. 3) calculations of $R_{rupture}$ for a single bubble with the threshold strain $\varepsilon_{th}$ = 0.7 agreed within 1% accuracy, confirming the validity of our selection of the computational parameters.

It is well known that the dynamics of cavitation bubbles produced in water and tissue (or tissue phantom) are quite different due to elasticity and viscosity of the latter material [24 – 26]. We have compared the shapes of two-bubble ensembles created in water and in gelatin using laser pulse energies adjusted to produce bubbles of the similar size in both media (Fig 5). The dissimilarity of the shape of the bubbles produced in gelatin and water can be ascribed primarily to the difference in focusing optics. While bubbles in water were created by water immersion objective, which minimized spherical aberration, no such objective was available for gelatin, which led to more elongated bubble shape. The difference in size between the two bubbles in gelatin is associated with slightly different angle of incidence of the two laser beams. In contrast to water, breakdown in gelatin leaves permanent cracks, thus every new laser shot was produced at new place, which made the adjustment of focusing geometry more difficult, compared to the experiment in water. Nonetheless, as demonstrated in Fig. 5, flattening of the proximal sides of the bubbles in both cases is quite similar, which suggests that the induced material displacements are similar as well. Based on this finding, we assume – for the rest of this section only – that the model of flow during expansion of the bubbles derived for water is applicable to gelatin as well.

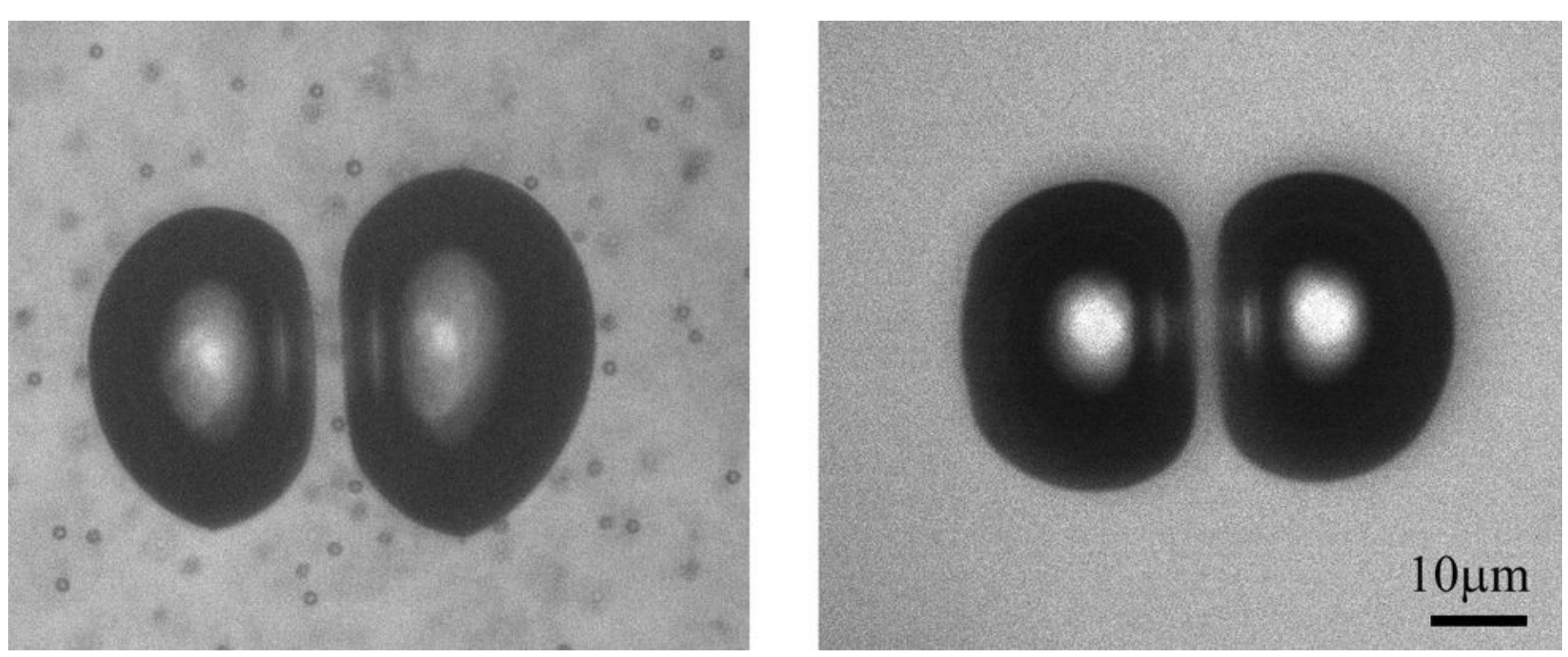

FIG. 5. Two simultaneously produced bubbles in gelatin (left) and water (right) at maximum expansion. The laser energy was adjusted to produce a single bubble of the same radius.

Two simultaneously expanding bubbles deform the tissue between them, and at sufficient proximity a continuous rupture zone can be created. For this to occur, the strain between the bubbles at the maximum expansion should exceed the threshold value. Assuming that the bubbles interaction in gelatin can be described similarly to that in water, the largest distance between the bubbles $L$, for which hydrodynamic interactions result in a continuous rupture zone with the threshold strain $\varepsilon_{th} = 0.7$ is $L$=2.37$R_{max}$. The radius of the rupture zone produced by a single bubble for the same value of the threshold strain $R_{rupture} = 0.7R_{max}$. The rupture zones created by simultaneous and sequential bubbles are shown in Fig. 6. Note that the interaction is localized between the bubbles, and practically does not affect the parts of the rupture zone at the opposite boundaries. Thus the entire length of the rupture zone produced by two simultaneously created bubbles is 3.77$R_{max}$ (2.37+2·0.7). Two bubbles applied sequentially create a cut of 2.8$R_{max}$ in length (4·0.7). Therefore, the total gain in the length of a cut is quite modest – by a factor of 1.35. If more than two bubbles are applied simultaneously, their effect is expected to be additive because of the localized influence of the neighboring bubble. Thus, the gain in the rupture zone produced by a large number of simultaneous bubbles would reach 1.69 (2.37/(2·0.7)). Both gain factors exhibit weak dependence on the threshold strain, and for $\varepsilon_{th} = 2$ they reach the value of 1.59 for two bubbles, and 2.17 for multiple bubbles. Trapping of the residual gas bubbles in tissue may further complicate focusing of subsequent pulses in sequential approach [24, 25]. Hence the efficiency of the sequential method derived above is an upper estimate.

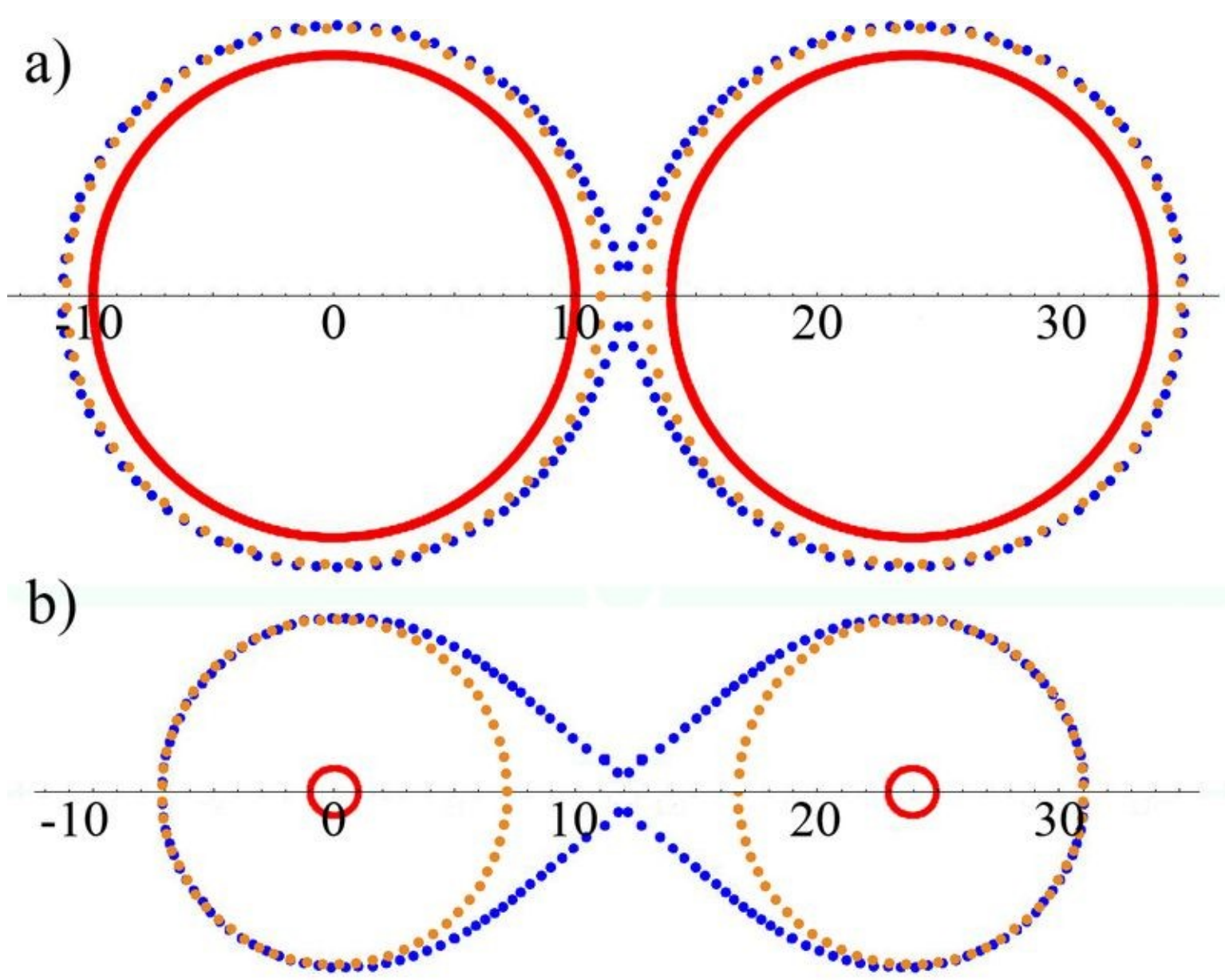

FIG. 6. (Color online) a) Rupture zone created by two bubbles at maximum expansion; b) Rupture zone after collapse of the bubbles. Solid red line – the initial bubble shape, orange (light) dotted line – rupture zone due to sequentially applied bubbles, blue (dark) dotted line – rupture zone created by two simultaneous bubbles.

### C. Experimental visualization of tissue deformation

Deformation of a transparent material can be imaged using embedded visible markers. A 1ms long time-integrated image shown in Fig. 7 demonstrates an example of the tracks left by 1μm polystyrene beads in gelatin, during the course of the bubbles expansion and collapse. Note that a bead exactly in-between the bubbles was not displaced, as expected from the symmetry. This approach allows for visualization of deformations in the materials of any complexity, as long as a sufficient number of visible markers can be embedded into it. However, in the current study we did not take this method beyond the illustration.

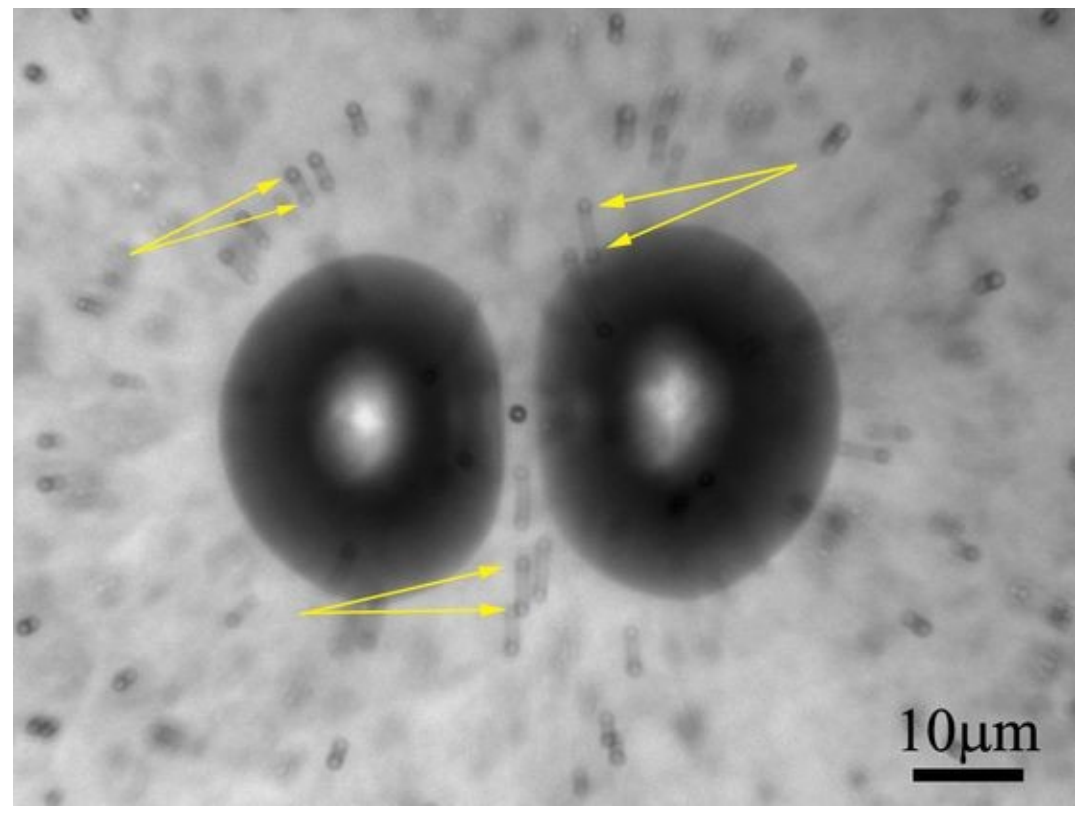

FIG. 7. (Color online) Displacements of 1μm polystyrene beads in gelatin resulting from the expansion of two cavitation bubbles. Arrow pairs indicate the initial and final positions of a few exemplary beads.

### D. Hydrodynamic interactions during collapse

From the considerations of symmetry, dynamics of the two simultaneous cavitation bubbles equal in size is equivalent to that of a single bubble next to a rigid boundary. Previous studies have shown that the presence of the boundary results in an asymmetric bubble collapse. During this stage the bubble drifts towards the wall and a high-speed jet is formed in the direction of the drift [27]. A similar behavior was observed in the current experiments with two bubbles, as long as their separation did not exceed ~$2.9R_{max}$, which corresponds to the gain in bubble separation of 1.54. The bubble attraction, accompanied by the jet formation and subsequent merger during the second expansion is shown in Fig. 8. Unlike in water, such behavior has not been observed in gelatin due to its relatively high viscosity and elasticity, therefore this effect is unlikely to be present in tissues of similar consistency. However, a thin layer of soft material immersed in a liquid, such as a membrane, located between the bubbles can in principle be ruptured by this process. A single bubble originating at small distance from a flexible membrane expands and pushes the membrane away, thereby decreasing the probability of rupture. In contrast to this, a membrane located between the two simultaneously created bubbles is "trapped" and is more likely to be ruptured either by the expanding bubbles, or by the jets produced during the bubble collapse.

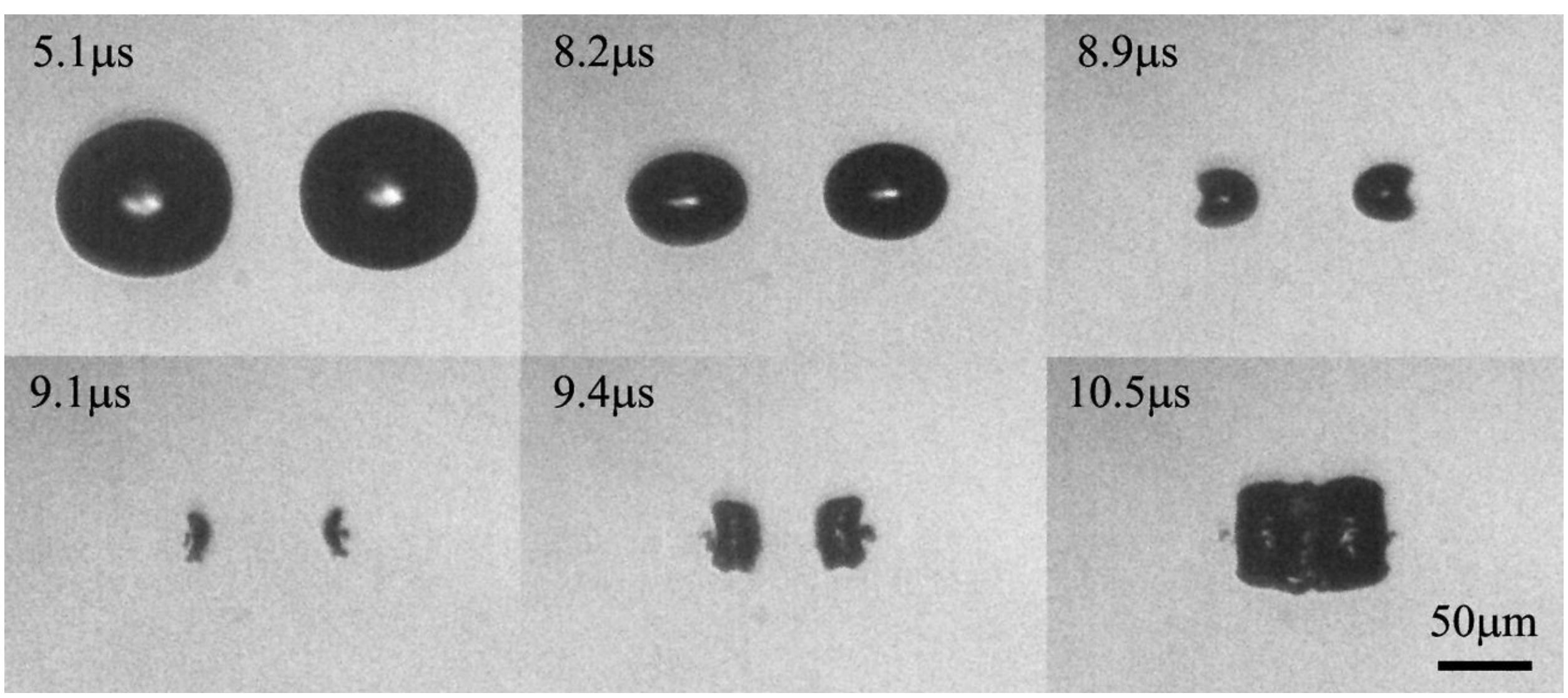

FIG. 8. Attraction and subsequent coalescence of cavitation bubbles in water during the collapse stage. Bubbles were separated by $2.9R_{max}$.

## V. CONCLUSIONS

An analytical model of the flow produced by two simultaneous cavitation bubbles has been developed and experimentally verified. The model allows for computing the displacements induced by the flow, and associated deformations in the medium. In addition, a method for direct measurement of such displacements in transparent material and a method for measurement of the threshold strain have been established. Using the analytical model with the experimentally obtained threshold strain we determined that two simultaneous bubbles in the material with a threshold strain of 0.7 increase the length of the rupture zone by a factor of 1.35, compared to the sequential approach. With multiple foci along the line of intended cut this enhancement factor can reach approximately 1.7. Bubbles collapsing in the materials with low viscosity (such as water) exhibit stronger interaction in the form of colliding jets, when the gap between their boundaries at the maximum expansion stage does not exceed 0.9 $R_{max}$. This interaction may extend the cutting zone by a factor 1.54, and can potentially be used for poration or cutting of the membranes trapped between the two foci.

## ACKNOWLEDGMENTS

The work of I.T, D.S. and D.P. was funded by the U.S. Air Force Office of Scientific Research (AFOSR) (contract FA9550-04-1-0075).

## APPENDIX A: DYNAMICS OF A SINGLE BUBBLE

The problem of a single cavitation bubble simplifies due to the spherical symmetry, so that all the unknowns depend only on the spherical radius, *r*, and time, *t*. The boundary conditions then become:

$$\left.\frac{\partial \Phi}{\partial r}\right|_{r=a} = \dot{a}\,; \tag{A1}$$

$$\left[\frac{\partial \Phi}{\partial t} + \frac{1}{2}\left(\frac{\partial \Phi}{\partial r}\right)^2\right]\Bigg|_{r=a} = -\frac{P(V) - P_\infty}{\rho}\,. \tag{A2}$$

It is natural to look for the solution $\Phi_{single}$ in the form:

$$\Phi_{single}(r,t) = \frac{A(t)}{r}\,, \tag{A3}$$

satisfying the Laplace equation. The function *A(t)* should be determined from the boundary conditions. Substituting the expression (A3) in the equations (A1) and (A2), we obtain:

$$A = -\dot{V}/3\,, \tag{A4}$$

$$\dot{A} + \frac{A^2}{2V} = -V^{1/3} g(V)\,, \tag{A5}$$

where $V=a^3(t)$ and $g(V)$ is defined by the Eq. (9). Introduction of (A4) to the eq. (A5) leads to the single equation for $V(t)$:

$$\ddot{V} - \frac{\dot{V}^2}{6V} - 3V^{1/3} g(V) = 0\,. \tag{A6}$$

By the standard procedure this autonomous equation can be reduced to the first order equation for $\dot{V}(t)$ whose solution gives *V* as an implicit function of time:

$$t = \int_{V_0}^{V(t)} \frac{dv}{\left(6v^{1/3}\int_{V_0}^{v} g(w)dw\right)^{1/2}}\,. \tag{A7}$$

Equation (8) for *A(t)* follows directly from (A4) and (A7). It is straightforward to verify that this solution coincides with the one derived by Rayleigh for $g(V) = -P_\infty/\rho$.

## APPENDIX B: FIRST ORDER SOLUTION FOR TWO BUBBLES

The normal to a surface defined by Eq. (11) can be written as:

$$\hat{n} = \frac{1}{\sqrt{1+\frac{f_\theta^2}{(a+f)^2}}}\left(\hat{e}_r - \frac{f_\theta}{a+f}\hat{e}_\theta\right) = \left(1-\frac{f_\theta^2}{2a^2}\right)\hat{e}_r - \left(\frac{f_\theta}{a} - \frac{ff_\theta}{a^2}\right)\hat{e}_\theta + ... \qquad \text{(B1)}$$

with the last expression valid to the second order in $f/a$, which represents the deviation of the actual bubble shape from the spherical one. Taking into account the axial symmetry, the velocity normal to the surface $v_n$ to the same second order can be written as:

$$v_n = \left(1-\frac{f_\theta^2}{2a^2}\right)\left(\dot{a}+\dot{f}+f_\theta\dot{\theta}\right) - \left(\frac{f_\theta}{a} - \frac{ff_\theta}{a^2}\right)(a+f)\dot{\theta} = \dot{a}+\dot{f} - \frac{\dot{a}f_\theta^2}{2a^2} + \ldots \qquad \text{(B2)}$$

Since the actual shape of the bubble boundary is unknown, we use the Taylor series representation around the unperturbed boundary and thus "shift" the boundary conditions to the spherical surface. In particular, the left-hand sides of the first and second boundary conditions become:

$$\left.\frac{\partial\Phi}{\partial\hat{n}}\right|_{r=a+f} = \left.\frac{\partial\Phi}{\partial\hat{n}}\right|_{r=a} + f\left(\frac{\partial}{\partial r}\left.\frac{\partial\Phi}{\partial\hat{n}}\right|\right)_{r=a} + \frac{f^2}{2}\left.\left(\frac{\partial^2}{\partial r^2}\frac{\partial\Phi}{\partial\hat{n}}\right)\right|_{r=a} + \ldots \qquad \text{(B3)}$$

$$\left.\left[\Phi_t + \frac{1}{2}(\nabla\Phi)^2\right]\right|_{r=a+f} = \left.\left[\Phi_t + \frac{1}{2}(\nabla\Phi)^2\right]\right|_{r=a} + f\left.\left[\Phi_{rt} + \frac{1}{2}\frac{\partial}{\partial r}(\nabla\Phi)^2\right]\right|_{r=a}$$

$$+ \frac{f^2}{2}\left.\left[\Phi_{rrt} + \frac{1}{2}\frac{\partial^2}{\partial r^2}(\nabla\Phi)^2\right]\right|_{r=a} + \ldots \qquad \text{(B4)}$$

Here and below the subscript denotes the differentiation with respect to the appropriate variable. Using the definition of normal derivative and the expression (B1) for the normal, we can rewrite eq. (B3) as:

$$\left.\frac{\partial\Phi}{\partial\hat{n}}\right|_{r=a+f} = \Phi_r\big|_{r=a} + \left.\left(-\frac{f_\theta}{a^2}\Phi_\theta + f\Phi_{rr}\right)\right|_{r=a}$$

$$+ \left.\left(-\frac{f_\theta^2}{2a^2}\Phi_r + \frac{2f_\theta f}{a^3}\Phi_\theta - \frac{f_\theta f}{a^2}\Phi_{r\theta} + \frac{f^2}{2}\Phi_{rrr}\right)\right|_{r=a} = H\Phi\big|_{r=a}, \qquad \text{(B5)}$$

where $H$ denotes the whole preceding linear differential operation. The right hand side of the second boundary condition also requires expansion, since it depends on the bubble volume $V$ and hence on $f$:

$$V = 2\pi\int_0^\pi \sin\theta d\theta \int_0^{a+f} r^2 dr = \frac{4\pi}{3}a^3 + 2\pi a^2\int_0^\pi f\sin\theta d\theta + 2\pi a\int_0^\pi f^2\sin\theta d\theta + \ldots \qquad \text{(B6)}$$

Denoting the three terms on the right as $V^{(0)}$, $V^{(1)}$, $V^{(2)}$, respectively, we then write the expansion for $g(V)$ as

$$g(V) = g(V^{(0)}) + g'(V^{(0)})V^{(1)} + \frac{1}{2}g''(V^{(0)})(V^{(1)})^2 + g'(V^{(0)})V^{(2)} + \ldots \qquad \text{(B7)}$$

Note that all the above calculations and formulas go for each of the two bubbles, which should be marked by the subscript 1 or 2, as the Cartesian and spherical coordinates $z_{1,2}$, $r_{1,2}$, $\theta_{1,2}$, etc. However, this subscript here and under similar circumstances below, is skipped.

The potential can be represented as a sum of the two potentials corresponding to the two independent bubbles and a small correction, namely:

$$\Phi(r,t) = \frac{A_1(t)}{r_1} + \frac{A_2(t)}{r_2} + \Phi^{(1)}(r_1,t). \qquad \text{(B8)}$$

We solve the equations at the first bubble boundary, skipping again the subscript 1, since it does not result in any ambiguity. The representation (B8) presumes that the bubble interaction acts as a perturbation, which is definitely true when each of the bubble radii is significantly smaller than the distance between them, $a/L<<1$. This assumption, however, turns out working well enough even in the case when the (identical) bubbles touch, and the above dimensionless parameter is equal to one half.

Plugging the expression (B8) and its derivatives into (B5) we obtain:

$$-\frac{A_1}{a^2}+\frac{A_2\cos\theta}{L^2}+\frac{A_2 f_\theta \sin\theta}{aL^2}+\frac{2A_1 f}{a^3}+\frac{A_1 f_\theta^2}{2a^4}-\frac{A_2 f_\theta^2\cos\theta}{2a^2L^2}-$$
$$\frac{A_2 f_\theta f\sin\theta}{a^2L^2}-\frac{3A_1 f^2}{a^4}+(H\Phi^{(1)})\Big|_{r=a}=\dot a+\dot f-\frac{\dot a f_\theta^2}{2a^2}. \qquad \text{(B9)}$$

The angle $\theta=\theta_1$ appears from the expression for $r_2$:

$$r_2=\sqrt{L^2+r_1^2-2r_1L\cos\theta_1}\ , \qquad \text{(B10)}$$

expanded properly in the above small parameter. The first terms on both sides cancel each other as the zeroth order solution [see (A4)], and so do the fifth term on the left and the last term on the right. Using the expansion (11) of $f$ and expressing $H$ according to its definition, we obtain:

$$\Phi_r^{(1)}\Big|_{r=a}=\left[\left(\frac{f_\theta^{(1)}}{a^2}\Phi_\theta^{(1)}-f^{(1)}\Phi_{rr}^{(1)}\right)\Big|_{r=a}+\dot f^{(1)}-\frac{A_2 f_\theta^{(1)}\sin\theta}{aL^2}-\frac{2A_1 f_\theta^{(1)}}{a^3}\right]$$
$$+\dot f^{(2)}-\frac{A_2\cos\theta}{L^2}-\frac{A_2 f_\theta^{(2)}\sin\theta}{aL^2}-\frac{2A_1 f^{(2)}}{a^3}+\frac{A_2 (f_\theta^{(1)})^2\cos\theta}{2a^2L^2}$$
$$+\frac{A_2 f^{(1)} f_\theta^{(1)}\sin\theta}{a^2L^2}+\frac{3A_1 (f^{(1)})^2}{a^4}. \qquad \text{(B11)}$$

Here only the terms in the square bracket are of the first order, the rest belongs to higher orders.
In a similar way we use expansion (B8) for the second boundary condition (B4), to find:

$$\frac{\dot A_1}{a}+\frac{\dot A_2}{L}+\frac{\dot A_2 a\cos\theta}{L^2}+\frac{A_1^2}{2a^4}-\frac{A_1A_2\cos\theta}{a^2L^2}+\left[\Phi_t^{(1)}-\frac{A_1}{a^2}\Phi_r^{(1)}+\frac{1}{2}(\nabla\Phi^{(1)})^2\right]\Big|_{r=a}$$
$$+f\left[-\frac{\dot A_1}{a^2}-\frac{2A_1^2}{a^5}+\Phi_{tr}^{(1)}+\frac{2A_1}{a^3}\Phi_r^{(1)}-\frac{A_1}{a^2}\Phi_{rr}^{(1)}\right]\Big|_{r=a}+f^2\left(\frac{\dot A_1}{a^3}+\frac{5A_1^2}{a^6}\right)$$
$$=-g(V^{(0)})-g'(V^{(0)})V^{(1)}-\left(\frac{1}{2}g''(V^{(0)})(V^{(1)})^2+g'(V^{(0)})V^{(2)}\right). \qquad \text{(B12)}$$

Introducing the expansion of $f$ (eq. 11) to the eq. (B12) and rearranging terms, we arrive at the equation:

$$\Phi_t^{(1)}\Big|_{r=a}=\left[\frac{A_1}{a^2}\Phi_r^{(1)}\Big|_{r=a}-\frac{\dot A_2}{L}+f^{(1)}\left(\frac{\dot A_1}{a^2}+\frac{2A_1^2}{a^5}\right)-g'\left(V^{(0)}\right)V^{(1)}\right]$$
$$-\frac{\dot A_2 a\cos\theta}{L^2}+\frac{A_1A_2\cos\theta}{a^2L^2}-\frac{1}{2}\left(\nabla\Phi^{(1)}\right)^2\Big|_{r=a}-f^{(1)}\left(\Phi_{tr}^{(1)}+\frac{2A_1}{a^3}\Phi_r^{(1)}-\frac{A_1}{a^2}\Phi_{rr}^{(1)}\right)\Big|_{r=a}$$
$$+f^{(2)}\left(\frac{\dot A_1}{a^2}+\frac{2A_1^2}{a^5}\right)-\left(f^{(1)}\right)^2\left(\frac{\dot A_1}{a^3}+\frac{5A_1^2}{a^6}\right)-\left(\frac{1}{2}g''(V^{(0)})(V^{(1)})^2+g'(V^{(0)})V^{(2)}\right), \qquad \text{(B13)}$$

where again all the first order terms are grouped in the square bracket. The resulting boundary conditions for the first order solution can be satisfied by setting $f^{(1)}\equiv 0$, when they reduce to

$$\Phi_r^{(1)}\Big|_{r=a}=0,\quad \Phi_t^{(1)}\Big|_{r=a}=-\frac{\dot A_2}{L}. \qquad \text{(B14)}$$

These are asymptotically satisfied if one takes $\Phi^{(1)}$ in the form:

$$\Phi^{(1)}=B_1(t)r_1\cos\theta_1+B_2(t)r_2\cos\theta_2 \qquad \text{(B15)}$$

with the functions $B_1(t)$ and $B_2(t)$ to be determined. Note that on the surface of the bubble 1, the angle $\theta_2$ is small, so, to the lowest order, its cosine is unity, sine is zero, and their derivatives in $r_1$ have the appropriate values as well; the higher order corrections to them are also available. This is used immediately below and in the Appendix C.

The second of the boundary conditions (B14) (for both bubbles) converts into the equation for $B_i$ (i=1,2):

$$\dot{B}_i = -\frac{\dot{A}_i}{L^2}\,, \tag{B16}$$

whose obvious integral with the initial conditions $A_i(0)=B_i(0)=0$ is $B_i=A_i/L^2$. Then the first condition (B14) is automatically valid, to this order, because $B_i=O((a/L)^2)$. This provides the answer for $\Phi^{(1)}$; being combined with the representation (B8), it allows for the total potential in the form

$$\Phi = A_1(t)\left(\frac{1}{r_1} - \frac{r_1\cos\theta_1}{L^2}\right) + A_2(t)\left(\frac{1}{r_2} - \frac{r_2\cos\theta_2}{L^2}\right) + \Phi^{(2)}\,, \tag{B17}$$

now with a correction of the second order. Finding it and the next one (all we actually need in this application) is algorithmically similar to the above calculations, although significantly more cumbersome; we describe it briefly in the next section.

## APPENDIX C: HIGHER ORDER CORRECTIONS FOR TWO BUBBLES

The results from the Appendix B can now be incorporated into the boundary conditions (B11) and (B13) to yield:

$$\Phi_r^{(2)}\Big|_{r=a} = \frac{A_1 - 2A_2}{L^2}\cos\theta + \dot{f}^{(2)} - \frac{2A_1 f^{(2)}}{a^3}\,, \tag{C1}$$

$$\Phi_t^{(2)}\Big|_{r=a} = \frac{\dot{A}_1 - 2\dot{A}_2}{L^2}a\cos\theta + \frac{\dot{A}_1 f^{(2)}}{a^2} + \frac{A_1\dot{f}^{(2)}}{a^2} - g'(V^{(0)})V^{(22)}\,, \tag{C2}$$

where $V^{(22)}$ is a non-zero part of $V^{(2)}$:

$$V^{(22)} = 2\pi a^2\int_0^{\pi} f^{(2)}\sin\theta d\theta\,. \tag{C3}$$

The structure of these conditions suggests a natural *Anzatz* for the second order correction to the bubble shape: $f^{(2)}(\theta, t)=\alpha(t)\cos\theta$, where $\alpha(t)$ is the function to be determined. Note that such representation nullifies $V^{(22)}$, according to its definition (C3). The boundary conditions then become:

$$\Phi_r^{(2)}\Big|_{r=a} = \left(\frac{A_1 - 2A_2}{L^2} + \dot{\alpha} - \frac{2A_1\alpha}{a^3}\right)\cos\theta\,; \tag{C4}$$

$$\Phi_t^{(2)}\Big|_{r=a} = \left(\frac{\dot{A}_1 - 2\dot{A}_2}{L^2}a + \frac{\dot{\alpha}A_1 + \alpha\dot{A}_1}{a^2}\right)\cos\theta\,. \tag{C5}$$

By substituting $\Phi^{(2)}$ in the form:

$$\Phi^{(2)} = \frac{C_1(t)}{r_1^2}\cos\theta_1 + \frac{C_2(t)}{r_2^2}\cos\theta_2 \tag{C6}$$

into (C4) and (C5) we immediately obtain equations (15) and (16), from which functions $\alpha(t)$, $C_1(t)$, and $C_2(t)$ can be determined (see the remark on the behavior of the j-th angle at the i-th bubble in Appendix B).

The third order approximation is derived in a similar way; however, more terms in Taylor series need to be accounted for. The normal to the boundary and the normal velocity are expressed to the third order as:

$$\hat{n} = \left(1 - \frac{f_\theta^2}{2a^2} + \frac{ff_\theta^2}{a^3}\right)\hat{e}_r - \left(\frac{f_\theta}{a} - \frac{f_\theta f}{a^2} + \frac{f_\theta f^2}{a^3} - \frac{f_\theta^3}{2a^3}\right)\hat{e}_\theta + \ldots \tag{C7}$$

$$v_n = \dot{a} + \dot{f} - \frac{\dot{a}f_\theta^2}{2a^2} + \left(-\frac{\dot{f}f_\theta^2}{2a^2} + \frac{\dot{a}ff_\theta^2}{a^3}\right) + \ldots \tag{C8}$$

The normal derivative of the potential is:

$$
\begin{aligned}
\left.\frac{\partial\Phi}{\partial\hat{n}}\right|_{r=a+f} &= \Phi_r\big|_{r=a} + \left.\left(-\frac{f_\theta}{a^2}\Phi_\theta + f\Phi_{rr}\right)\right|_{r=a} \\
&+ \left.\left(-\frac{f_\theta^2}{2a^2}\Phi_r + \frac{2f_\theta f}{a^3}\Phi_\theta - \frac{f_\theta f}{a^2}\Phi_{r\theta} + \frac{f^2}{2}\Phi_{rrr}\right)\right|_{r=a} \\
&+ \left(\frac{ff_\theta^2}{a^3}\Phi_r - \frac{3f_\theta f^2}{a^4}\Phi_\theta + \frac{f_\theta^3}{2a^4}\Phi_\theta - \frac{ff_\theta^2}{2a^2}\Phi_{rr}\right. \\
&\left.\left. + \frac{2f_\theta f^2}{a^3}\Phi_{r\theta} - \frac{f_\theta f^2}{2a^2}\Phi_{rr\theta} + \frac{f^3}{6}\Phi_{rrrr}\right)\right|_{r=a} \\
&= G\Phi\big|_{r=a},
\end{aligned}
\tag{C9}
$$

where the operator $G$ is a shorthand notation for the middle expression in (C9). To obtain the correct expansion of volume $V$, a new term, referred to as $V^{(3)}$, has to be added to (B6):

$$
V = 2\pi\int_0^\pi \sin\theta d\theta \int_0^{a+f} r^2 dr = \frac{4\pi}{3}a^3 + 2\pi a^2\int_0^\pi f\sin\theta d\theta + 2\pi a\int_0^\pi f^2\sin\theta d\theta + \frac{2\pi}{3}\int_0^\pi f^3\sin\theta d\theta. \tag{C10}
$$

The expansion for $g(V)$ then is:

$$
\begin{aligned}
g(V) &= g(V^{(0)}) + g'(V^{(0)})V^{(1)} + \frac{1}{2}g''(V^{(0)})(V^{(1)})^2 + g'(V^{(0)})V^{(2)} \\
&+ g'(V^{(0)})V^{(3)} + g''(V^{(0)})V^{(1)}V^{(2)} + \frac{1}{6}g'''(V^{(0)})(V^{(1)})^3.
\end{aligned}
\tag{C11}
$$

The left hand side of the second boundary condition is expanded in the following series:

$$
\begin{aligned}
\left.\left[\Phi_t + \frac{1}{2}(\nabla\Phi)^2\right]\right|_{r=a+f} &= \left.\left[\Phi_t + \frac{1}{2}(\nabla\Phi)^2\right]\right|_{r=a} + f\left.\left[\Phi_{rt} + \frac{1}{2}\frac{\partial}{\partial r}(\nabla\Phi)^2\right]\right|_{r=a} \\
&+ \frac{f^2}{2}\left.\left[\Phi_{rrt} + \frac{1}{2}\frac{\partial^2}{\partial r^2}(\nabla\Phi)^2\right]\right|_{r=a} \\
&+ \frac{f^3}{6}\left.\left[\Phi_{rrrt} + \frac{1}{2}\frac{\partial^3}{\partial r^3}(\nabla\Phi)^2\right]\right|_{r=a}.
\end{aligned}
\tag{C12}
$$

We already know the solution up to the second order, therefore we can express the potential and boundary shape as:

$$
\Phi = A_1(t)\left(\frac{1}{r_1} - \frac{r_1\cos\theta_1}{L^2}\right) + A_2(t)\left(\frac{1}{r_2} - \frac{r_2\cos\theta_2}{L^2}\right) + \frac{C_1(t)\cos\theta_1}{r_1^2} + \frac{C_2(t)\cos\theta_2}{r_2^2} + \Phi^{(3)}; \tag{C13}
$$

$$
R(\theta,t) = a(t) + \alpha(t)\cos\theta + f^{(3)}. \tag{C14}
$$

These expansions are introduced into (C9) and (C12) to obtain, after rearranging the terms:

$$
\Phi_r^{(3)}\Big|_{r=a} = \dot{f}^{(3)} - \frac{2A_2 a}{L^3}P_2(\cos\theta) - \frac{2A_1 f^{(3)}}{a^3}; \tag{C15}
$$

$$
\Phi_t^{(3)}\Big|_{r=a} = -\frac{\dot{A}_2 a^2}{L^3}P_2(\cos\theta) + \frac{f^{(3)}\dot{A}_1}{a^2} + \frac{\dot{f}^{(3)}A_1}{a^2} - g'(V^{(0)})V^{(3)}. \tag{C16}
$$

These boundary conditions can be satisfied by setting $f^{(3)}=\beta(t)P_2(\cos\theta)$:

$$
\Phi_r^{(3)}\Big|_{r=a} = \left(\dot{\beta} - \frac{2A_2 a}{L^3} - \frac{2A_1\beta}{a^3}\right)P_2(\cos\theta); \tag{C17}
$$

$$\Phi_t^{(3)}\Big|_{r=a} = \left( -\frac{\dot{A}_2 a^2}{L^3} + \frac{\beta \dot{A}_1}{a^2} + \frac{\dot{\beta} A_1}{a^2} \right) P_2(\cos\theta) . \tag{C18}$$

In deriving (C18) we took into account that our choice of $f^{(3)}$ leads to $V^{(3)}\equiv 0$. The third-order approximation is thus found in the form:

$$\Phi^{(3)} = \frac{D_1(t)P_2(\cos\theta_1)}{r_1^{\;3}} + \frac{D_2(t)P_2(\cos\theta_2)}{r_2^{\;3}} . \tag{C19}$$

Equations (19) and (20) that define the functions $\beta(t)$ and $D_1(t)=D_2(t)$ follow from substituting of the representation (C19) into the boundary conditions (C17) and (C18). Similarly to the previous order correction, the second term in the sum (C19) turns out to be of higher order, at the first bubble boundary.